\documentclass[]{spie}  

\newcommand{\ignore}[1]{}

\usepackage{amsmath,amsfonts,amssymb}
\usepackage{graphicx}
\usepackage{float}
\usepackage[colorlinks=true, allcolors=blue]{hyperref}

\title{Low noise flux estimate and data quality control monitoring in EUCLID-NISP cosmological survey}

\author[a]{Bogna Kubik}
\author[a]{Remi Barbier}
\author[a]{Peter Calabria}
\author[a]{Alain Castera}
\author[a]{Eric Chabanat}
\author[a]{Florence Charlieu}
\author[b]{Jean-Claude Clemens}
\author[b]{Anne Ealet}
\author[a]{Sylvain Ferriol}
\author[b]{William Gillard}
\author[c]{Thierry Maciaszek}
\author[d]{Eric Prieto}
\author[a]{Florent Schirra}
\author[b]{Aurelia Secroun}
\author[b]{Benoit Serra}
\author[a]{Gerard Smadja}
\author[b]{Antre Tilquin}
\author[b]{Julien Zoubian}

\affil[a]{Institut de Physique Nucl\'eaire de Lyon, 4 rue Enrico Fermi Villeurbanne 69622, France}
\affil[b]{Centre de Physique des Particules de Marseille, 163 avenue de Luminy 13288 Marseille, France}
\affil[c]{Ctr. National d'\'Etudes Spatiales, France}
\affil[d]{Aix Marseille Universit\'e, CNRS, LAM UMR 7326, 13388, Marseille, France}
\authorinfo{Further author information: (Send correspondence to B. Kubik)\\B. Kubik: E-mail: bkubik@ipnl.in2p3.fr, Telephone: +33 4 72 44 58 52}

\begin{document} 
\maketitle

\begin{abstract}
Euclid mission is designed to understand the dark sector of the universe. Precise redshift measurements are provided by H2RG detectors. We propose an unbiased method of fitting the flux with Poisson distributed and correlated data, which has an analytic solution and provides a reliable quality factor - fundamental features to ensure the goals of the mission. We compare our method to other techniques of signal estimation and illustrate the anomaly detection on the flight-like detectors. Although our discussion is focused on Euclid NISP instrument, much of what is discussed will be of interest to any mission using similar near-infrared sensors.
\end{abstract}


\section{INTRODUCTION}
\label{sec:intro}
In this work we focus on the signal extraction method and the data quality control in the near infrared (NIR) channel of Euclid satellite. A high accuracy of the on board signal estimate, a rigorous control of the error and fit quality as well as the calibration with sub-percent precision are fundamental to ensure the scientific goals of the mission. For this purpose we have introduced a new statistical estimator of the signal and an associated quality factor of the fit to the data produced by the NIR detector in the Euclid focal plane. 

Euclid NIR detectors are read out in the multiple accumulated sampling (MACC) mode. The accumulating signal is sampled up the ramp as a function of time and the multiple reads are averaged within groups. The averaging of the digitized frames within groups reduces the readout noise of each pixel. The signal is then estimated in each pixel by fitting a line to the averaged groups as a function of time. 

Euclid's telemetry limitations do not allow to transfer to the ground all of the averaged groups for the subsequent processing. For each exposure the signal accumulated in the focal plane, sampled over more than $60\times10^6$ pixels, is fitted on board and the image of the estimated signals is sent to the ground. The applied algorithm of signal fit is subject to the CPU limitations and must be analytic. The anomalies appearing during the integration, such as cosmic ray hits, nonlinear response or electronic instabilities, any of them provoking an inconsistent fit result, have to be detected with the image of quality factors. The latter is transferred to the ground for each exposure in addition to the signal layer.

Usually an equally weighted least square fit is applied to estimate the signal. Although this procedure has an analytic solution, the derivation of the corresponding quality factor requires a second pass over all the groups digitized up the ramp per pixel, which is impossible within the CPU limitations of Euclid. Moreover, the equally weighted least square fit neglects the correlations between the digitized frames and treats the samples as Gaussian distributed with constant variance. It was noticed in (\cite{doi:10.1117/1.JATIS.1.3.038001}) that the optimal calculation of the flux, in terms of signal to noise ratio, should use all the information contained in the covariance matrix. In this case, however, explicit analytic expressions for the parameters cannot be obtained and the parameters must be computed using numerical methods which are too demanding in terms of accessible memory and computing time.

As a very stringent error budget is required in the near-infrared channel of Euclid we have introduced an adapted statistical estimator of the signal and the associated fit quality factor from Poisson distributed and correlated data. Both the estimator and the quality factor have an analytic form and thus can be easily implemented in the digital processing unit on board. The flux bias is kept under control in the range of representative science fluxes and the associated error is by 6\% lower than the commonly used formula derived in (\cite{2007PASP..119..768R,10.1086/656514}) in the context of an equally least square fit. 

In this work, we test the method of fitting the flux and detecting anomalies in nondestructive read exposures, proposed and tested with Monte Carlo simulations in (\cite{Kubik:2016pasp}), using data taken with engineering grade Euclid-like H2RG detectors. The paper is structured as follows. In section \ref{sec:method} we present the nondestructive multi-accumulated readout scheme of the H2RG detectors and we the method of fitting the nondestructive read ramps and the definition of the quality factor. Next, in section \ref{sec:results} we present the results of measurements performed on the engineering grade Euclid-like detectors. We compare signal and error computed using the method proposed in (\cite{Kubik:2016pasp}) to the commonly used least square fit. Afterwards, we illustrate how the quality factor is sensitive to various kind of anomalies that can appear during the exposure. The anomalies include the nonlinear response and saturation, cosmic ray hits and electronic induced jumps in the pixel level 
during integration. 

Although our discussion is focused on the Euclid NIR array, we anticipate that much of what is discussed should be of interest to any mission using similar near-infrared sensors.

\section{METHOD}
\label{sec:method}

\subsection{H2RG multi accumulated readout principle and importance of working in the differential signal mode}
\label{subsec:nondestructive_readout}
\begin{figure}[H]
    \begin{center}
        \begin{tabular}{cc}
            \includegraphics[scale=0.25]{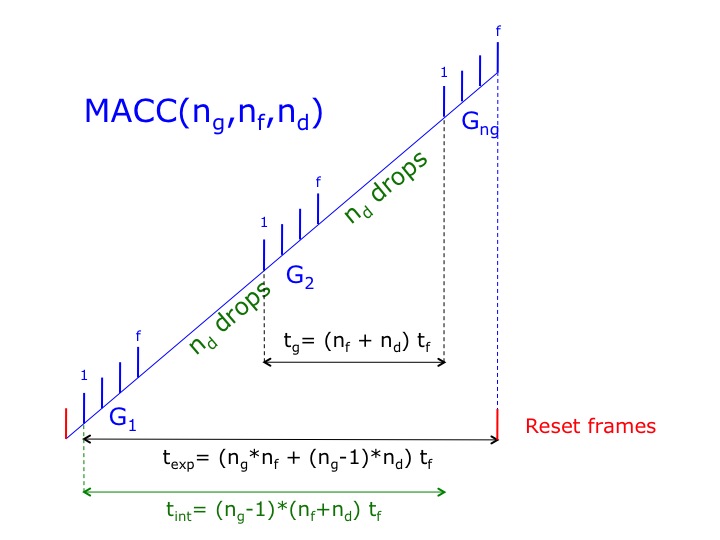}
        \end{tabular}
    \end{center}
    \caption{\footnotesize Multiple Accumulated sampling MACC($n_g$, $n_f$, $n_d$) with $n_g$ - number of equally spaced groups sampled up the ramp, $n_f$ - number of frames per group and $n_d$ - number of dropped frames between two successive groups.\label{fig:readout_modes}}
\end{figure}

For science observations NIR detectors, in general, acquire up the ramp (UTR) sampled data at a constant frame cadence. A frame $S_i$ is the unit of data that results from sequentially clocking through and reading out a rectangular area of pixels during a time $t_f$. As in the UTR mode individual frames suffer from a high readout noise, it is common to average the frames within groups $G_k$. This readout pattern is called a multiple accumulated sampling and frequently a common abbreviation MACC($n_g$, $n_f$, $n_d$) is in use where $n_g$ is the number of equally spaced groups sampled up the ramp, $n_f$ is the number of frames per group and $n_d$ is the number of dropped frames between two successive groups (see Figure \ref{fig:readout_modes}).
Before fitting the flux, the frames within groups are averaged. The signal in group $G_k$ after averaging $n_f$ frames is equal to
\begin{equation}
    G_k = \frac{1}{n_f}\sum_{i=1}^{n_f} S_i^{(k)}.
\end{equation}
The advantage of the coadding procedure is the reduction of the Gaussian distributed pixel readout noise $\sigma_R$, assumed here to be uncorrelated from frame to frame, down to the effective group read noise $\sigma_{RG}$ defined by
\begin{equation}
    \sigma_{RG}^2 = \frac{\sigma_R^2}{n_f}.
\end{equation}
Therefore, the larger the number of frames per group, the lower the readout noise associated to each group after coadding.

The correlations between data points are intrinsic to the non-destructive readouts of the integrating arrays. If one works with the groups $G_i$, the Poisson and readout noise correlate all the points $(G_k, G_l)$ (\cite{doi:10.1117/1.JATIS.1.3.038001}) implying a use of numerical fitting procedure.

A large part of Poisson noise contributions to the correlations between groups is removed if one works with the differences of the groups $\Delta G_k = G_{k+1} - G_k$. The variance associated with $\Delta G_k = G_{k+1} - G_k$ is equal to (\cite{doi:10.1117/1.JATIS.1.3.038001})
\begin{equation}\label{eq:Dkk}
   D_{kk}(\Delta G_k;g) = (1+\alpha)\frac{g}{f_e} + 2\sigma_{RG}^2\, ,
\end{equation}
where the Poisson noise correlations are encoded in the coefficient $\alpha$ 
\begin{equation}\label{eq:coefs_a_b_c}
    \alpha  = \frac{1-n_f^2}{3n_f(n_f+n_d)}\, .
\end{equation}
$D_{kk}(\Delta G_k;g)$ includes the effective electronic readout noise of a pixel $\sigma_{RG}$, the flux Poisson noise ($\propto\frac{g}{f_e}$) and the Poisson noise correlations arising during the coadding procedure ($\propto \alpha\frac{g}{f_e}$). We explicit the dependence on the conversion gain $f_e$ [$e^-$/ADU] which converts the arbitrary digital units (ADU), recorded by the readout electronics, into electrons. In this way Eq. (\ref{eq:Dkk}) and the formulas in the following sections apply directly to the raw pixel data not converted into physical units, i.e. the inter-group flux $g$  is measured in ADU per unit time $t_g = (n_f+n_d)t_f$, and the electronic readout noise $\sigma_R$ is given in ADU. $D_{kk}(\Delta G_k;g)$ is the error on the measurement $\Delta G_{k}$ and depends on the inter-group flux $g$.

The remaining non-null correlations between group differences $(k, k\pm1)$
\begin{equation}\label{eq:Dkl}
    D_{k,k+1}(\Delta G_k;g) = \frac{(n_f^2-1)}{6n_f(n_f+n_d)}\frac{g}{f_e}-\frac{\sigma_R^2}{n_f}\, ,
\end{equation}
which take into account the correlation of the Poisson and readout noise between two {\it consecutive} group differences, can be safely neglected at fluxes higher than 5 $e^-$/sec as they represent not more than 4\% of the diagonal terms for the typical readout noise value $\sigma_R = 10$ $e^-$ in the MACC(15,16,11) exposure mode. In the method presented in (\cite{Kubik:2016pasp}) and in the following sections the correlations $D_{k, k+1}$ are neglected for all the fluxes.

\subsection{Signal estimate definition}
\label{subsec:fitdef}
The model that describes the measurements of nondestructive readouts from infrared integrated arrays is defined by the likelihood $\mathcal{L}$ (\cite{Kubik:2016pasp})
\begin{equation}\label{eq:likelihood_general}
    \mathcal{L} = \prod_{i=1}^{n_g-1} \frac{1}{\sqrt{2\pi \sigma_{eff}^2(\Delta G_i, g) }}\exp{\left(-\frac{(\Delta G_i - g)^2 }{2 \sigma_{eff}^2(\Delta G_i, g)}\right)},
\end{equation}
where the total error squared on the measured differences $\Delta G_{k}$
\begin{equation}\label{eq:sig_eff}
    \sigma_{eff}^2(\Delta G_k; g) = D_{kk}(\Delta G_k;g)\, .
\end{equation}

The signal estimator $\hat{g}$ given by first derivative of $\left(-2\times\log\mathcal{L}\right)$ is equal to
\begin{equation}\label{eq:lflux_estimator}
    \hat{g} = \frac{1+\alpha}{2f_e}\left[ \sqrt{1 + \frac{4 f_e^2 \sum_{i=1}^{n_g-1} (\Delta G_i + \beta)^2 }{(n_g-1)(1+\alpha)^2}} - 1 \right] - \beta\, ,
\end{equation}
where 
\begin{equation}\label{eq:coefs_a_b_c}
    \beta = \frac{ 2\sigma_R^2 f_e }{ n_f(1+\alpha) }\, .
\end{equation}
Importantly, $\hat{g}$ has an analytic solution and thus can be easily implemented in the on board electronics. The model assumes that the detector response should be linear over its entire well depth, thermally stable, and not subject to any time dependent effects that vary during integrations or readouts. Any of these effects should be detected by the quality factor defined in the next section.

\subsection{Quality factor definition}
\label{subsec:qfdef}
The compatibility of the data with the straight line fit is tested by the quality factor $QF$ defined as 
\begin{equation}\label{eq:QF_definition_as_fcn_hatg}
    QF = \frac{2 f_e}{(1+\alpha)}[(n_g-1)\hat{g_x} - (G_n - G_1)]\, ,
\end{equation}
where the pseudo-flux estimator $\hat{g_x}$
\begin{equation}\label{eq:gx}
    \hat{g_x} = \sqrt{\sum_{i=1}^{n_g-1}\frac{(\Delta G_i + \beta )^2}{ n_g - 1 }} - \beta
\end{equation}
is the value of $g$ in the minimum of $QF$. The quality factor verifies whether the nondestructive readouts are not subject to any anomaly such as cosmic ray hits, RTS noise, electronic jumps or nonlinearity and saturation. In the absence of anomalies $QF$ follows the well known $\chi^2_{\textrm{th}}(x;n)$ distribution with $n$ degrees of freedom defined by
\begin{equation}\label{eq:chi2_distribution_def}
    \chi^2_{\textrm{th}}(x;n) =  \frac{(x^2)^{ \frac{n}{2}-1 } \exp^{ -\frac{x^2}{2} } }{ 2^{\frac{n}{2}}\Gamma( \frac{n}{2} ) }
\end{equation}
The mean of the $\chi^2_{\textrm{th}}(x;n)$ distribution is equal to $n$ - the number of degrees of freedom, which in our case is equal to $n_g-2$ and is independent of the flux value, the number of coadded frames $n_f$, and the number of drops $n_d$. We normalize the $\chi^2_{\textrm{th}}(x;n)$ distribution by dividing it by the number of degrees of freedom, therefore the distribution of the normalized quality factor is centered on 1. 

In the following section we show the normalized quality factor obtained with real data. We show that the distributions are centered on unity for a large range of fluxes. Furthermore, we give examples of pixels that are affected by anomalies such as cosmic ray hits, RTS noise, nonlinearity or saturation, making the estimated flux $\hat{g}$ unreliable. These anomalies can be detected on the basis of a value of the quality factor in a given exposure. In this paper, we do not intend to give the exact threshold values of $QF$ above which the pixels should be masked. The calibration of the $QF$ thresholds per pixel should be addressed in a future work.

\section{RESULTS}
\label{sec:results}
\subsection{Preliminary measurements of the detector properties}
\label{subsec:sR_fe}
We test the new fitting method on engineering grade Euclid H2RG $\lambda_{c} = 2.3\,\mu$m detectors operated at 90K in our test facilities. The readout noise $\sigma_R$ per pixel, entering the Eqs. (\ref{eq:lflux_estimator}) and (\ref{eq:QF_definition_as_fcn_hatg}) was computed on 19 dark exposures in UTR(800)$\equiv$MACC(800,1,0) mode. Each exposure was first corrected with reference pixels scheme $c_{3}(64,4)$ described in (\cite{doi:10.1117/12.2055071}) and the readout noise was computed on each ramp. The median value of the readout noises obtained in the 19 exposures was then computed to eliminate any spurious events that could potentially deteriorate the result of a single exposure.

\begin{figure}[H]
    \begin{center}
        \begin{tabular}{c}
            \includegraphics[height=6cm]{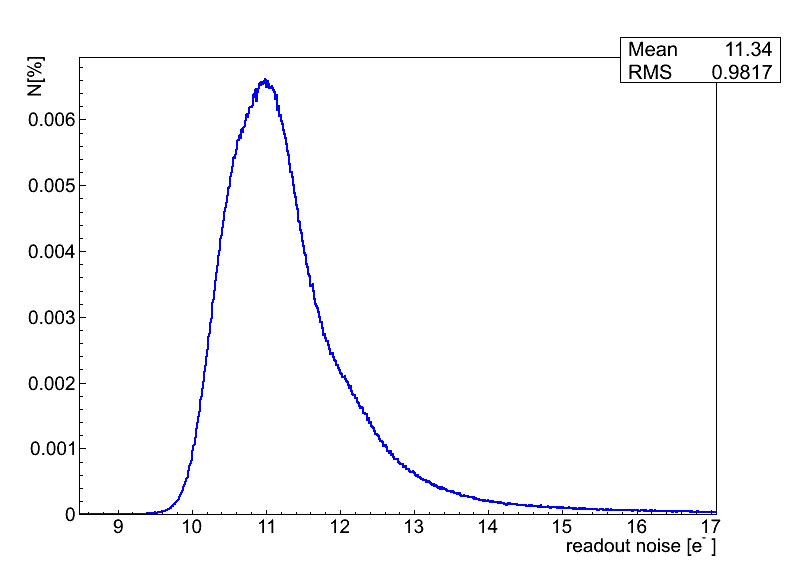}\\
            \includegraphics[height=6cm]{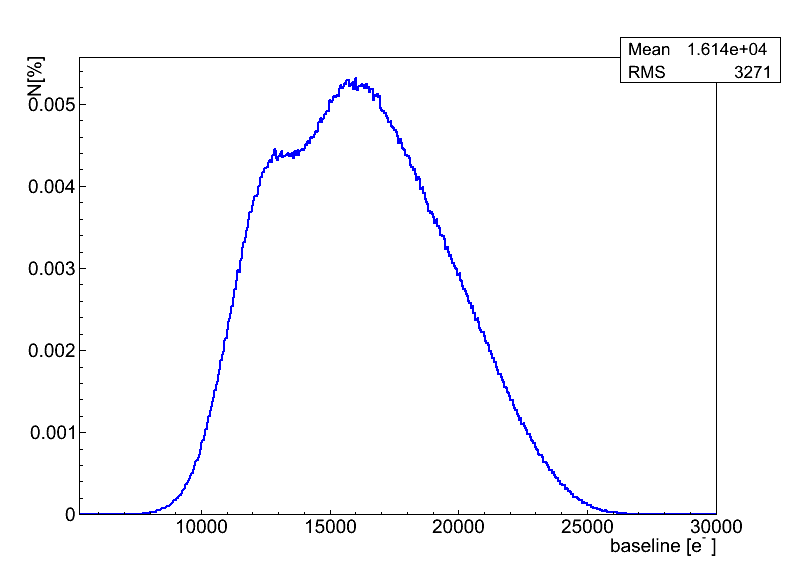}
            \includegraphics[height=6cm]{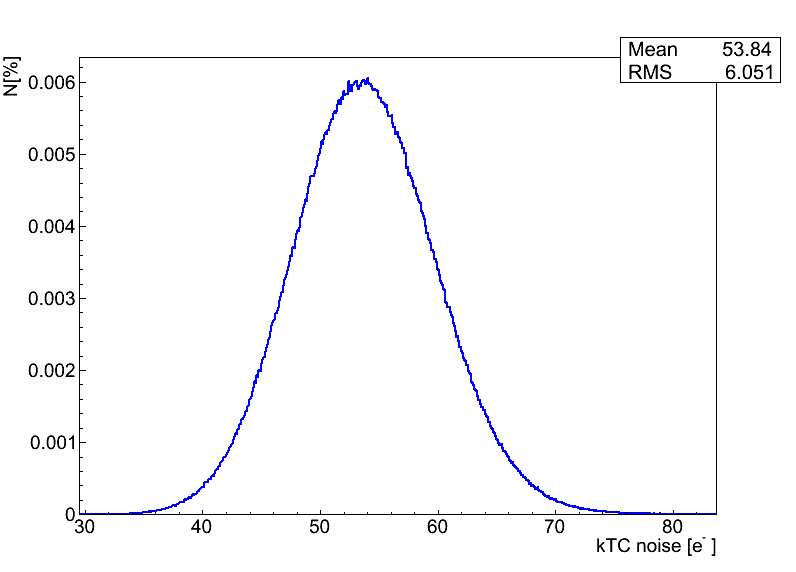}
        \end{tabular}
    \end{center}
   \caption[Distributions]
   {\label{fig:sR_baseline_distribution} Distributions of the readout noise (upper panel) baseline value (bottom left panel) and the associated kTC noise (bottom right panel) as measured over 40 exposures in dark conditions.}
\end{figure}

In upper panel of Figure \ref{fig:sR_baseline_distribution} we show the obtained distribution of the readout noise with the average value of 11 $e^{-}$ and a 10\% dispersion around the mean. We use the average conversion gain $f_e = $ 1.32 $e^-$/ADU, computed using the Photon Transfer Curve method on flat field exposures for each pixel. We also measured the average baseline value by taking the average image of the first frame after reset over 40 exposures in dark conditions. The baseline distribution and the associated kTC noise (dispersion around the mean baseline value) are shown in bottom panel of Figure \ref{fig:sR_baseline_distribution}. 

\subsection{Comparison of the fitting method to the least square fit algorithm}
\label{subsec:compareLSF}
\begin{figure}[H]
    \begin{center}
        \begin{tabular}{c}
            \includegraphics[height=6cm]{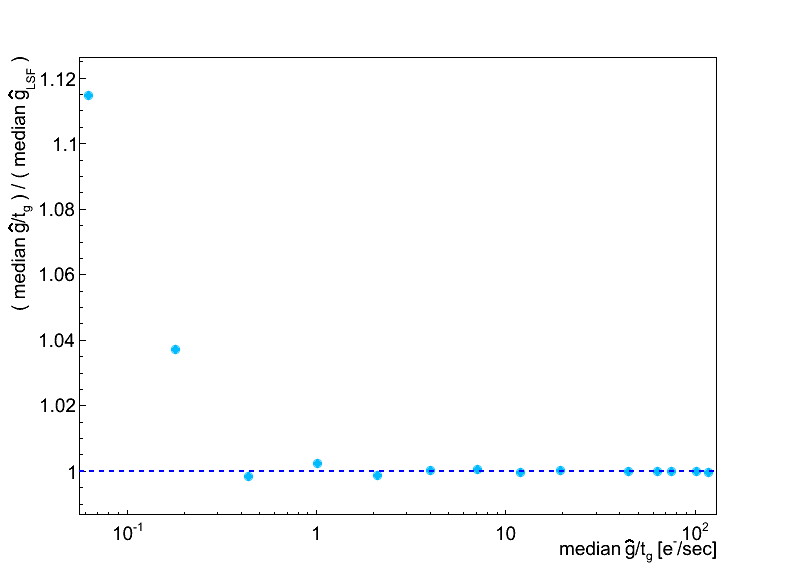}
            \includegraphics[height=6cm]{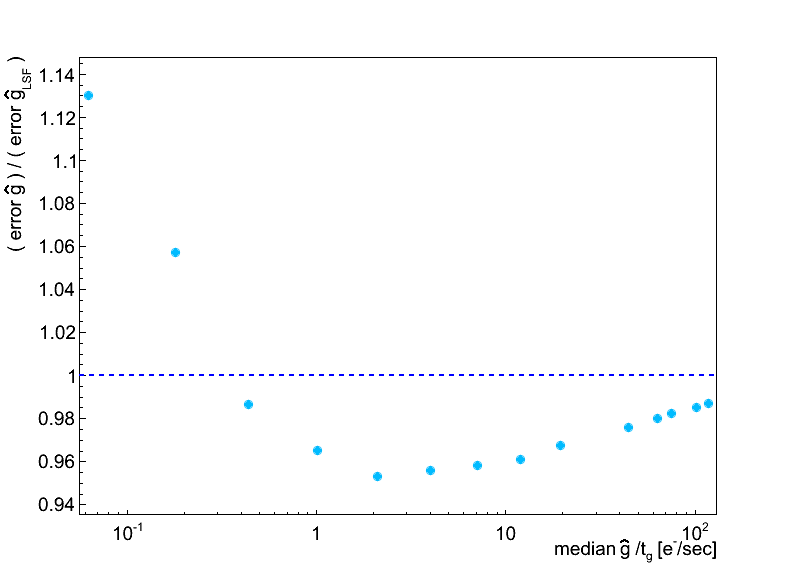}
        \end{tabular}
    \end{center}
   \caption[compareLSF]
   {\label{fig:compareLSF} Comparison of the method proposed in (\cite{Kubik:2016pasp}) with the least square algorithm. In left panel we plot the ration of the flux computed using Eq. (\ref{eq:lflux_estimator}) to the LSF result. In right panel we plot the ratio of the noise, computed as the spatial dispersion around the mean flux value on flat field exposures.}
\end{figure}

We compare the results of our fitting method to the commonly used least square fit (LSF) algorithm - a method that neglects any correlation of the sampled points and their Poisson distribution. 
In the left panel of Figure \ref{fig:compareLSF} we show the ratio of the signal computed using Eq. (\ref{eq:lflux_estimator}) to the LSF result. The agreement between the two methods is very good for all the tested fluxes down to 0.2 $e^{-}$/sec. Below this flux value fitting function Eq. (\ref{eq:lflux_estimator}) becomes sensitive to the readout noise correlations which results in a deviation from the LSF fit result. In the right panel of Figure \ref{fig:compareLSF} we plot the ratio of the noise associated to the two methods. The noise at each flux level was computed as the spatial median dispersion around the median flux value over the array. This plot confirms that at low fluxes ($<$ 0.2 $e^{-}$/sec) the presented method becomes sensitive to the readout noise correlations neglected in the fitting function. Moreover, it proves that the error of $\hat{g}$ is lower by 5 to 2\%, depending on the flux value, than the noise associated to the LSF algorithm. This confirms the predictions based on the 
Monte Carlo simulations presented in (\cite{Kubik:2016pasp}) and translates directly into the quality of scientific data and in the increased figure of merit of the survey. 

\subsection{Sensitivity of the quality factor to nonlinearity}
\label{subsec:NLsensitivity}

\begin{figure}[H]
    \begin{center}
        \begin{tabular}{c}
            \includegraphics[height=6cm]{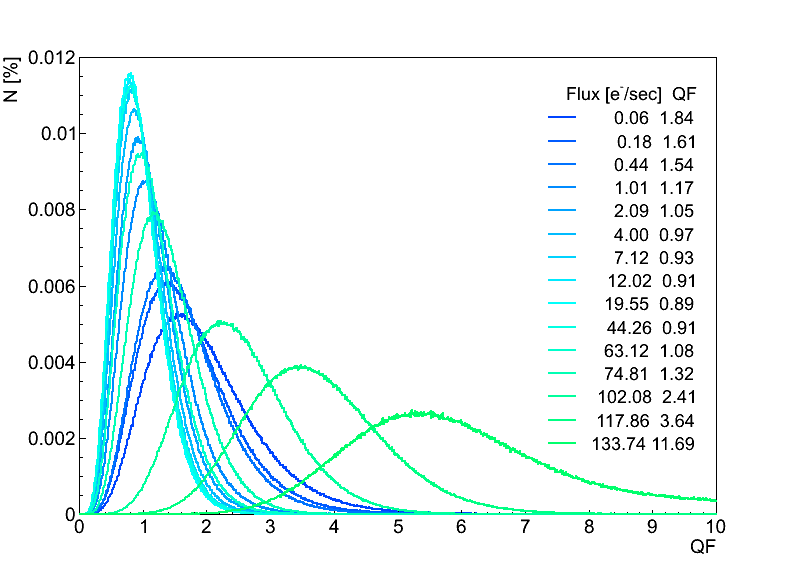}
            \includegraphics[height=6cm]{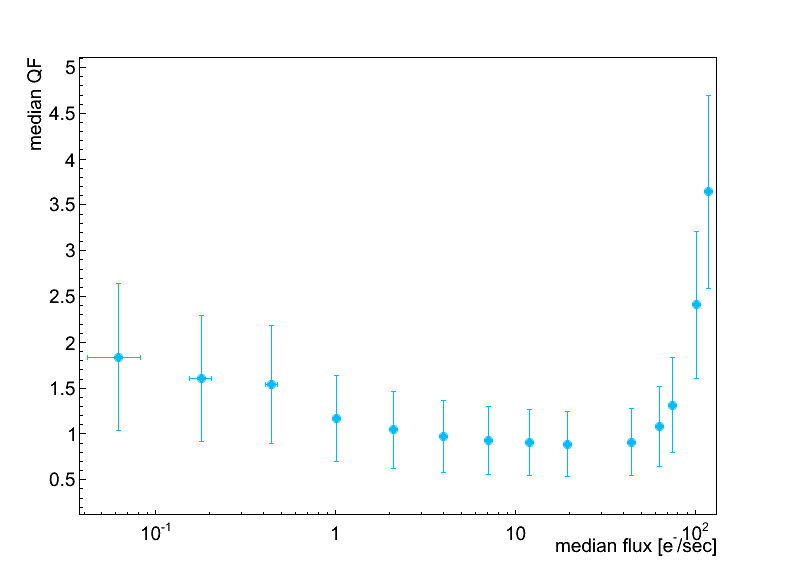}\\
            \includegraphics[height=6cm]{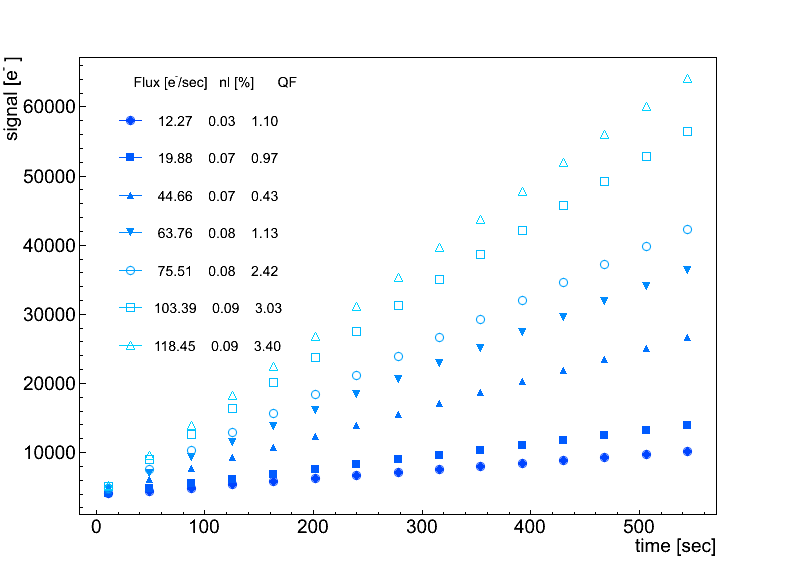}
        \end{tabular}
    \end{center}
   \caption[$QF$ distributions for various fluxes]
   {\label{fig:qf_distributions_all_fluxes} In the upper left panel we plot the distributions of normalized $QF$ obtained on real data. In the legend we report the median flux and median quality factors computed on the array on $2040\times 2040$ pixels each. In the upper right panel we plot the median $QF$ as function of the median flux per exposure. The error bars indicate the normalized median absolute deviation (nmad) of the distributions. In the bottom panel we show the integrations with incearsing flux for one pixel with mildly nonlinear behavior.}
\end{figure}

The near infrared CMOS arrays are inherently nonlinear devices. Due to this nonlinear behavior the differences $\Delta G_i$ measured later in an exposure will be smaller than those measured early in the exposure. Both the signal estimator $\hat{g}$ and the quality factor $QF$ are sensitive to this effect. We do not address here the issue of how to correct for the nonlinearity but rather show how the quality factor reacts to the nonlinear response of the pixels. To measure this effect we took flat field exposures with 15 values of fluxes increasing from 0.05 to 130 $e^{-}$/sec. All the exposures were taken in UTR(400), coadded to MACC(15,16,11) and corrected with reference pixels scheme $c_{3}(64,4)$ described in (\cite{doi:10.1117/12.2055071}). In the upper left panel of Figure \ref{fig:qf_distributions_all_fluxes} we plot the distributions of the computed quality factors at increasing fluxes. The median flux and median quality factor per exposure (medians computed on the arrays of $2040\times 2040$ 
pixels) are reported in the legend of the figure. In the upper right panel we plot the median $QF$ as function of the median flux. 

We observe that for the increasing flux, and therefore for the increasing deviation from the linear response, the quality factors increase. This is expected, as we try to fit a straight line, using Eq. (\ref{eq:lflux_estimator}) to the nonlinear exposure. On the other hand, for fluxes lower than 1 $e^{-}$/sec the median $QF$ is higher than 1. This is due to the fact the in Eq. (\ref{eq:lflux_estimator}) the correlations between the consecutive group differences are not taken into account. The correlations are negligible at higher fluxes but in the low flux regime the readout noise becomes comparable with the Poisson noise and the signal fit starts to be sensitive to its contributions. This effect was noticed in (\cite{Kubik:2016pasp}) with the  Monte Carlo simulation and is now recovered with the real data. Both effects should be calibrated by pixel and corrected before applying the quality flag threshold to mask other anomalies. 

In the bottom panel of Figure \ref{fig:qf_distributions_all_fluxes} we plot the cumulative signal as a function of time in one of one pixel (4,4) which has a mildly nonlinear response. The deviation from nonlinearity, reported in the middle column in the figure's legend, varies between 3\% and 9\%. The associated quality factor varies in the range between 1 and 3. This deviation from linear response will not be detected by the quality factor if the $QF$ threshold is put too high, nevertheless the estimated flux will have to be corrected for the nonlinearity effects using calibration data.

\subsection{Sensitivity of the quality factor to saturation}
\label{subsec:Saturationsensitivity}

\begin{figure}[H]
    \begin{center}
        \begin{tabular}{c}
            \includegraphics[height=6cm]{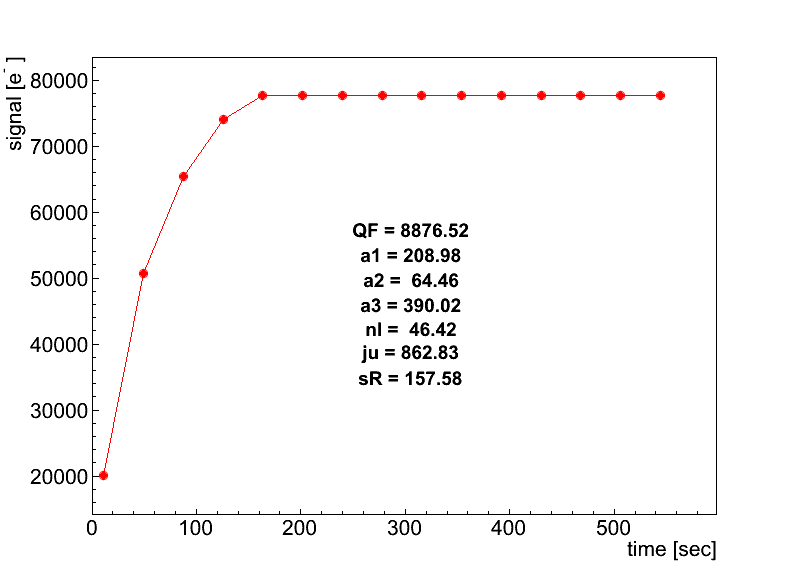}
            \includegraphics[height=6cm]{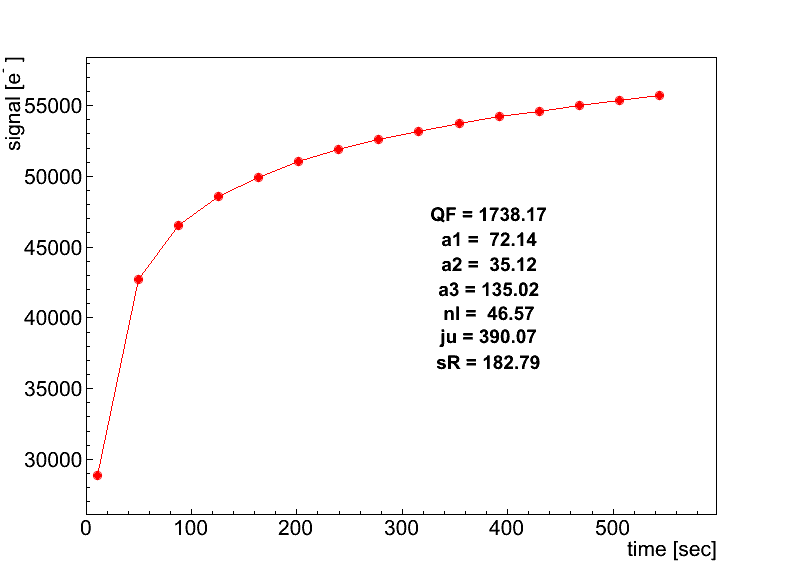}
        \end{tabular}
    \end{center}
   \caption[$QF$ on saturation]
   {\label{fig:qf_SAT} Quality factor on saturating ramps.}
\end{figure}

Some pixels reach the saturation level before the end of the exposure. This can be due to either a high value of the dark current, high pedestal level or high quantum efficiency as compared with the median value on the integrating array. Independently of the nature of the saturation this kind of anomalous behavior is readily detected by the quality factor as illustrated in Figure \ref{fig:qf_SAT}. In the left panel the signal accumulated in the pixel saturates the ADC and the measured quality factor is equal to 8876. This pixel saturated the ADC even in dark conditions in less than 100 seconds with the dark current of about 900 $e^{-}$/sec and a quality factor of 8316 in MACC(15,16,11). 
In the right panel of Figure \ref{fig:qf_SAT} the saturation appears more likely in the pixel diode and is also easily detected by $QF = 1738$. The dark current of this pixel is about 400 $e^-$/sec. In both examples the pixels' readout noise ($\sigma_R = 157,\,182$ respectively) is high, compared to the average readout noise of the array $\langle \sigma_R \rangle = 11$ $e^{-}$. 

In the pixels which saturate nearly at the end of the exposure, the quality factor will be lower but we have seen that the $QF$ values are typically higher than 500 in those cases. The exact $QF$ threshold for detecting the saturation should be studied in details depending on the intrinsic pixel properties such as the dark current, the readout noise and the pedestal level.

\subsection{Sensitivity of the quality factor to cosmic rays}
\label{subsec:CRsensitivity}

\begin{figure}[H]
    \begin{center}
        \begin{tabular}{c}
            \includegraphics[height=6cm]{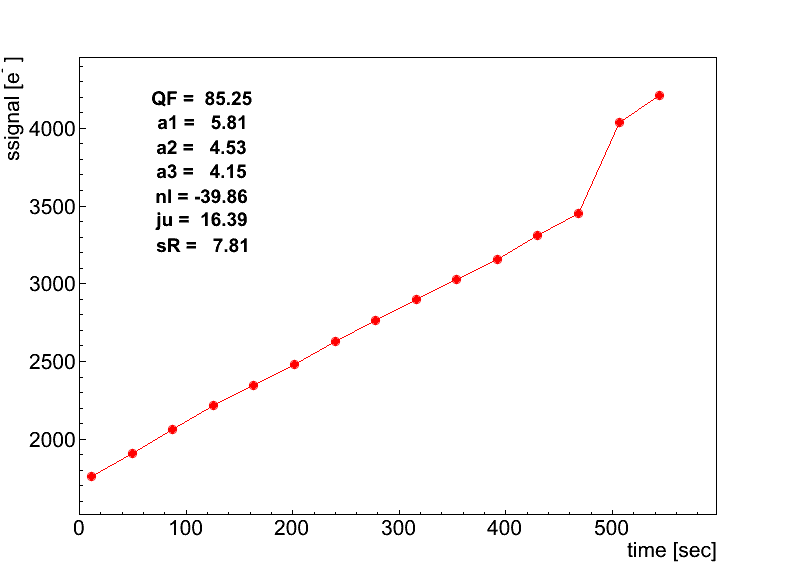}
            \includegraphics[height=6cm]{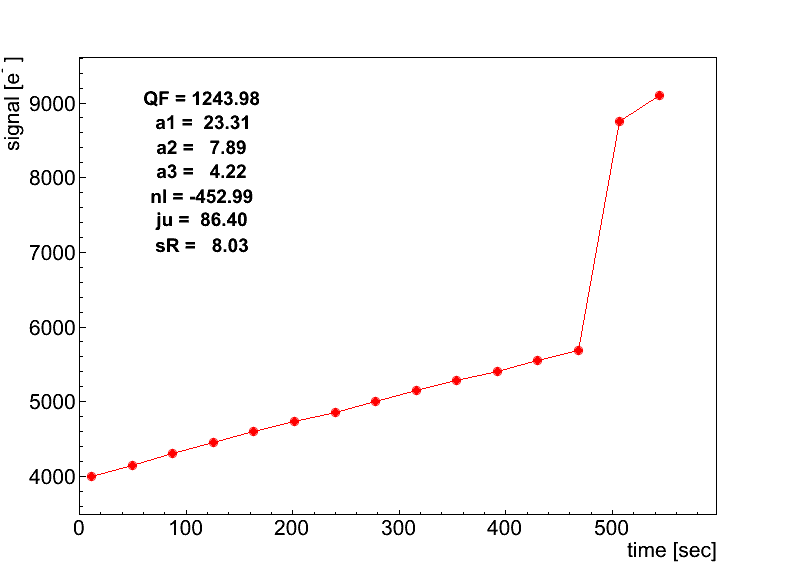}\\
            \includegraphics[height=6cm]{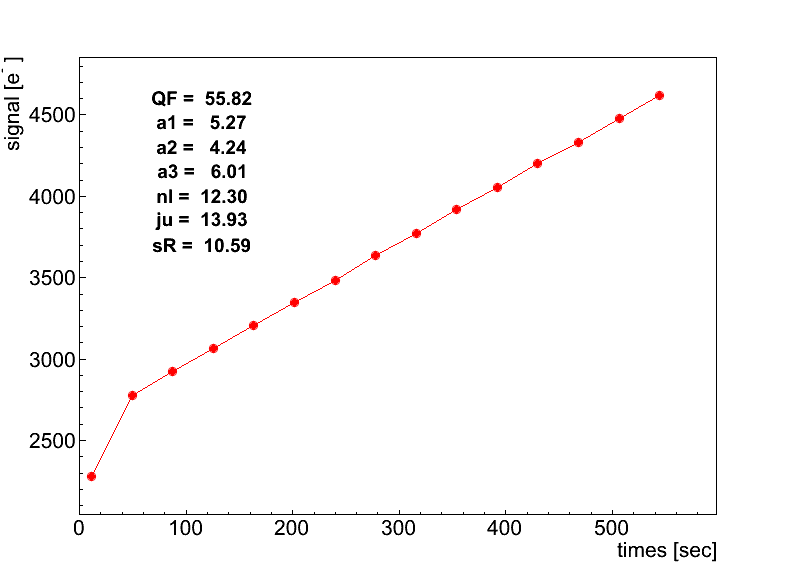}
            \includegraphics[height=6cm]{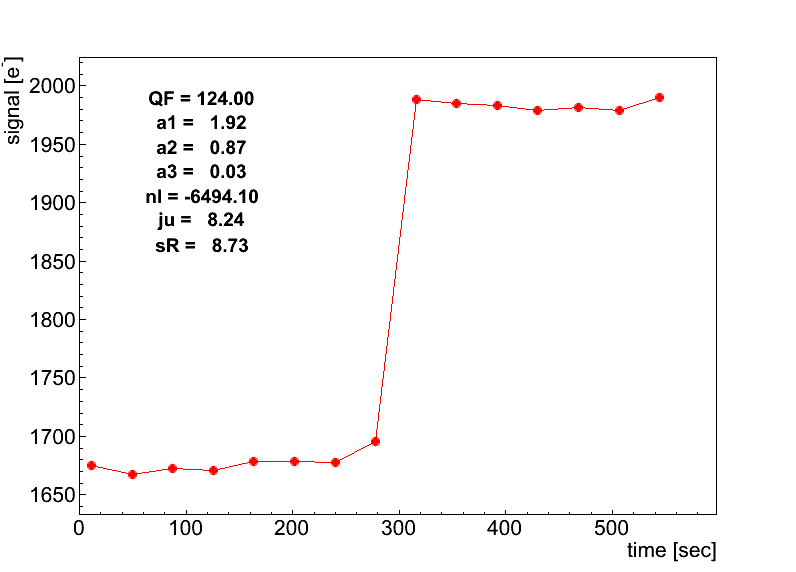}
        \end{tabular}
    \end{center}
   \caption[$QF$ on CR]
   {\label{fig:qf_CR} Examples of the quality factors on ramps affected by the cosmic ray hits.}
\end{figure}

Correcting for cosmic rays (CR) in nondestructive read ramps is not a new problem (\cite{Fixsen:2000ke,Gordon:2005ur,1538-3873-123-908-1237}). There are a lot of options for CR detection when the ground-based processing is applied to the ramps. However, if the ramps are processed in orbit, the applied algorithms can not be as complex as those usually applied on ground. We use the quality factor defined in Eq. (\ref{eq:QF_definition_as_fcn_hatg}) to detect the cosmic ray hits. As illustrated in Figure \ref{fig:qf_CR}, the exposures affected by the CR have typically a high value of $QF$. 

\begin{figure}[H]
    \begin{center}
        \begin{tabular}{c}
            \includegraphics[height=6cm]{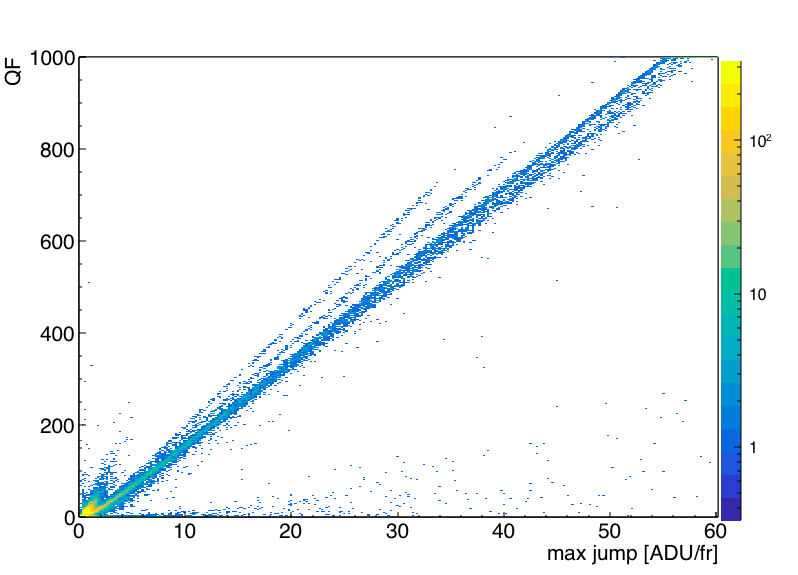}
        \end{tabular}
    \end{center}
   \caption[correlation $QF$ and CR hit amplitude]
   {\label{fig:qf_CR_correlation} Correlation between the quality factor and the amplitude of the cosmic ray hit ($\gamma$ rays emitted by the Fe${}^{55}$ source) in [ADU/fr] detected in dark exposures in ramps in MACC(15,16,11). The rapid increase of the $QF$ with the cosmic ray hit amplitude allows a detection of cosmic ray events during integrations. (This plot was done with data taken in the N\'eel Institute in Grenoble, witht the radioactive source Fe${}^{55}$. The authors would like to thank all the people involved during the measurements for their effort.)}
\end{figure}

The value of $QF$ depends on the amount of charge created in the pixel by the comic ray. In the upper left(right) panel the cosmic ray deposes around 600(3250) $e^{-}$ respectively. The associated quality factors are 85 and 1243. The higher the CR charge deposit, the higher the quality factor. The net correlation between the CR charge is further illustrated in Figure \ref{fig:qf_CR_correlation} where we plot the value of $QF$ as function of the cosmic ray charge deposit.

Moreover, the value of $QF$ does not depend of the hit position in the exposure. In the upper left and lower left panels of Figure \ref{fig:qf_CR} the charge deposits in the pixel are comparable (490 and 600 $e^{-}$  respectively). The corresponding $QF$s are equal to 55 and 85 and the small increase in the $QF$ value is due to the increase in the charge deposit. The lack of sensitivity of $QF$ to the hit position comes from the fact that we fit the differences between the groups and not the groups themselves. Therefore the CRs hits can be equally well rejected independently of the hit position in the exposure. One exception is the hit that appears in the first group. In this case the ramp offset will be shifted to a higher value but no jump will be seen in the ramp. The detailed sensitivity to the cosmic ray events as a function of the underlying flux and the pixel properties (readout noise, conversion factor) should be evaluated in a future work.

\subsection{Sensitivity of the quality factor to electronic anomalies}
\label{subsec:RTSsensitivity}
\begin{figure}[H]
    \begin{center}
        \begin{tabular}{c}
            \includegraphics[height=6cm]{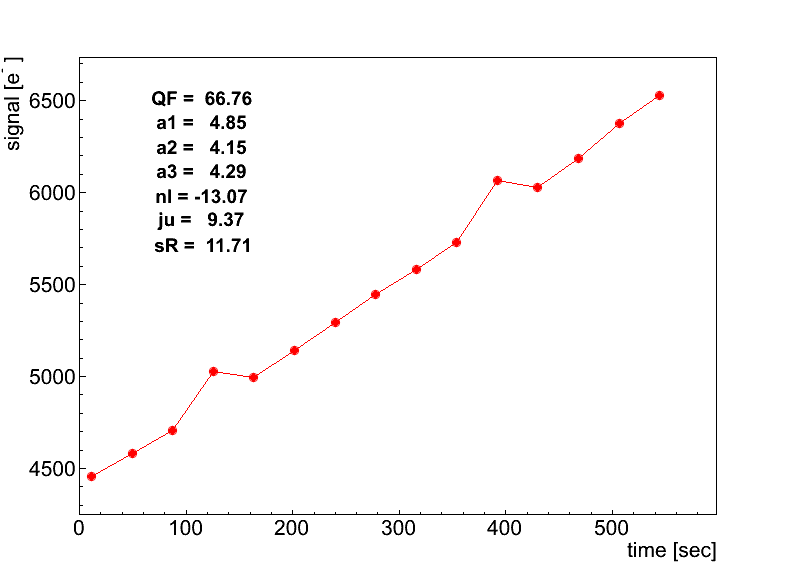}
            \includegraphics[height=6cm]{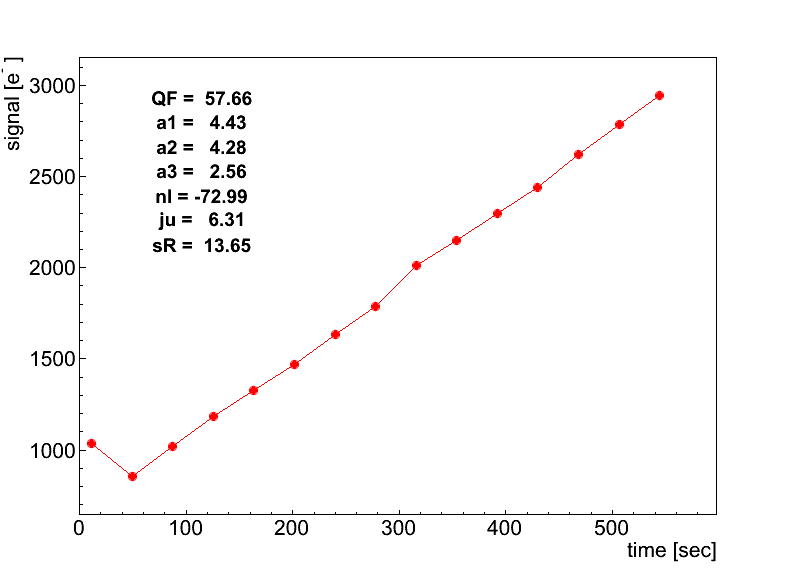}\\
            \includegraphics[height=6cm]{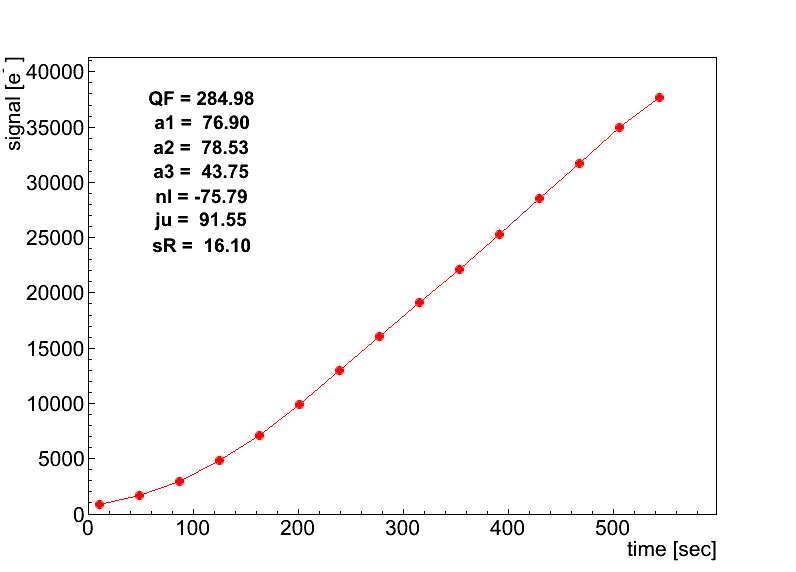}
            \includegraphics[height=6cm]{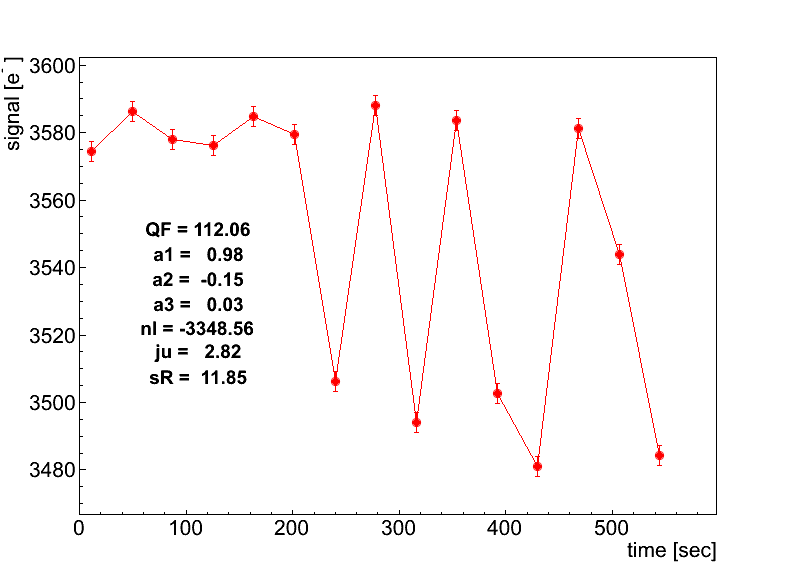}
        \end{tabular}
    \end{center}
   \caption[$QF$ on RTS]
   {\label{fig:qf_RTS} Examples of the quality factors on ramps with RTS noise and electronic jumps.}
\end{figure}

Finally, the quality factor is sensitive to the electronic instabilities and time variable effects during the exposures. In Figure \ref{fig:qf_RTS} we show the examples of pixels that are subject to such anomalies. These include single jumps during the exposures (upper left and right panels), "inverse" nonlinearity (bottom left panel) or a high RTS noise (bottom right panel). Each of these effects provokes an increase in the $QF$ value and can therefore be detected on the ground.

\subsection{Correlations of the quality factor with pixel properties}
\label{subsec:QFcorrelations}
\begin{figure}[H]
    \begin{center}
        \begin{tabular}{c}
            \includegraphics[height=6cm]{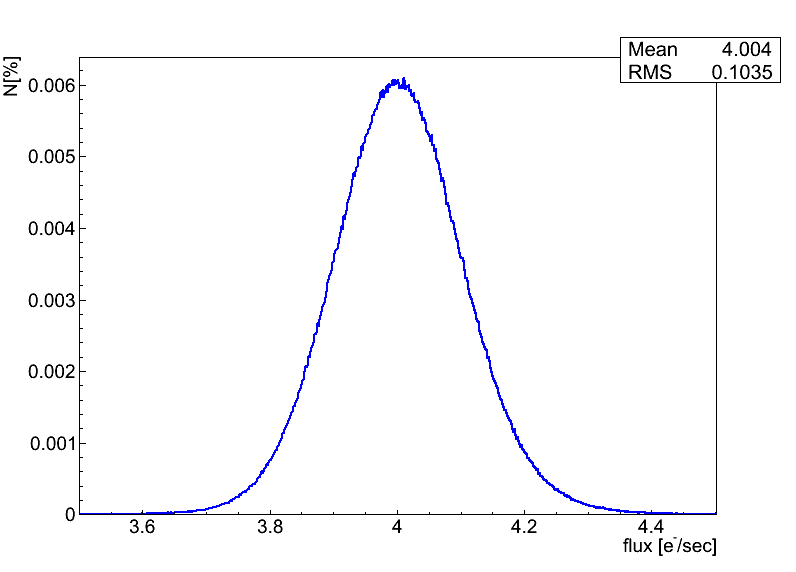}
            \includegraphics[height=6cm]{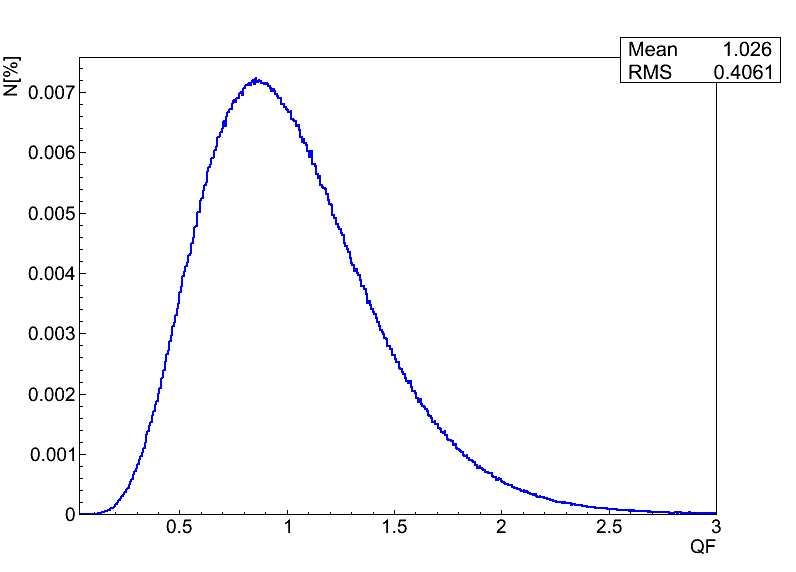}
        \end{tabular}
    \end{center}
   \caption[Distributions]
   {\label{fig:QF_dark_distributions} Distributions of the measured dark current (left panel) and the associated quality factor (right panel) obtained on one flat field exposure in MACC(15,16,11) mode.}
\end{figure}

\begin{figure}[H]
    \begin{center}
        \begin{tabular}{c}
            \includegraphics[height=6cm]{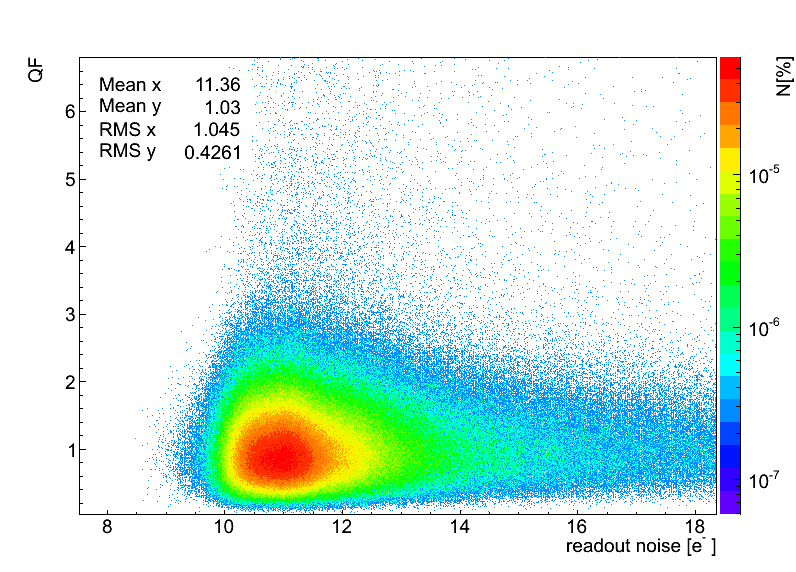}\\
            \includegraphics[height=6cm]{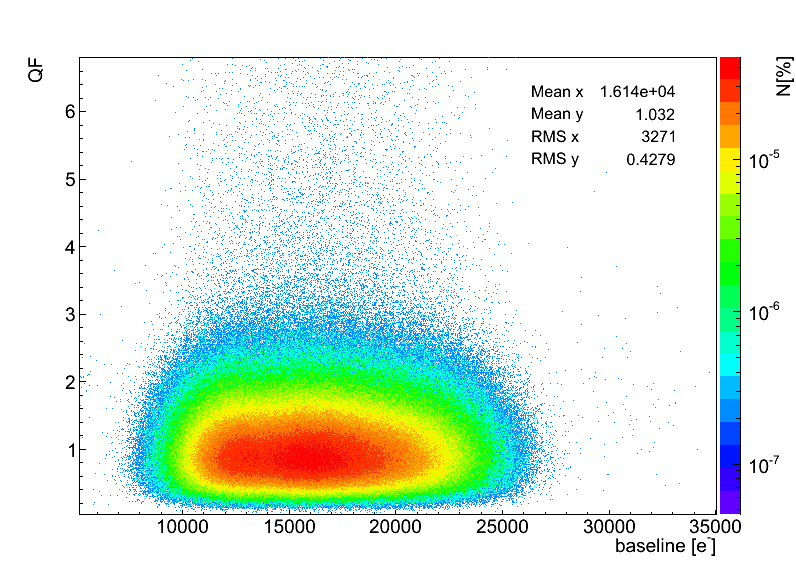}
            \includegraphics[height=6cm]{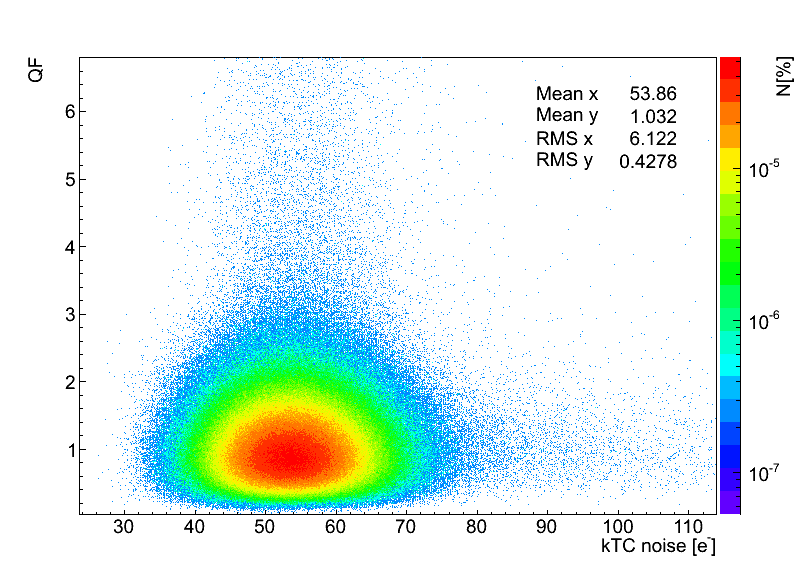}
        \end{tabular}
    \end{center}
   \caption[$QF$ correlations]
   {\label{fig:QF_correlations} Correlations between the quality factor and the readout noise (upper panel), baseline value (bottom left) and kTC noise value (bottom right). The quality factors were computed on one flat field exposure in MACC(15,16,11) mode. The readout noise, baseline and kTC nois were computed independently on the dark exposures as described in section \ref{subsec:sR_fe}.}
\end{figure}

Finally we verify whether there are any correlations between the quality factors with the readout noise, baseline, kTC noise and flux. We apply Eqs. (\ref{eq:lflux_estimator}) and (\ref{eq:QF_definition_as_fcn_hatg}) to one exposure in MACC(15,16,11) to compute the fluxes and the quality factors. The obtained distributions are plotted in Figure \ref{fig:QF_dark_distributions}. We are in the range where the ramps are linear and the flux is sufficiently high that the readout noise correlations are negligible. The quality factor distribution is centered on 1, indicating that the majority of pixels do not suffer from any anomaly. 
In Figure \ref{fig:QF_correlations} we plot the correlations of the quality factor and the pixels properties such as readout noise, baseline and the kTC noise. We observe that the quality factor is not correlated to the baseline, the kTC noise and readout noise values at the flux of 4 $e^{-}$/sec. Thus the $QF$ can be hardly used to detect pixels with too high baseline or excessive kTC noise or readout noise in low or medium flux conditions.

\section{SUMMARY}
\label{sec:summary}
In this work, we tested the method of fitting the flux and detecting anomalies in nondestructive read exposures, proposed and tested with Monte Carlo simulations in (\cite{Kubik:2016pasp}), using data taken with engineering grade Euclid-like H2RG detectors. The integrating arrays with the cutoff wavelength $\lambda_{c} = 2.3\,\mu$ were operated at 90K in our test facilities. The pixel properties entering into the equations for the flux and quality factor are the readout noise and the conversion gain. We computed the readout noise per pixel using 19 dark exposures in UTR(800). The mean value of the readout noise over the array is equal to 11.34 [$e^{-}$]. The average conversion gain $f_e = 1.32$ [$e^{-}$/ADU] was evaluated using the photon transfer curve. 

First, we have confirmed with real data that the two methods, the one presented in (\cite{Kubik:2016pasp}) and the unweighted least square fit, give the same signal value. Moreover, we have confirmed that the error on the flux estimator introduced (\cite{Kubik:2016pasp}) is lower by 5-2\% than the noise derived in \cite{2007PASP..119..768R,10.1086/656514} for fluxes higher than 2 $e^-$/sec. This translates directly into a higher figure of merit of the survey.

Next, we have computed the quality factor on exposures with increasing fluxes, from 0.05 up to 130 $e^{-}$/sec. At fluxes lower than 1 $e^{-}$/sec the quality factors are higher than the expected value of 1 because of the readout noise correlations between the consecutive groups. These correlations were neglected in the proposed method, but become important at very low flux levels where the Poisson noise associated to the signal is comparable to the contributions of the electronic readout noise. This effect was noticed in (\cite{Kubik:2016pasp}) in Monte Carlo simulations and is now confirmed with real data. As Euclid will not observe at fluxes lower than $1$ $e^{-}$/sec, this will not be an issue. However, in general case, one should correct for this effect while looking at the data.

In the range between 1 and 50 $e^{-}$/sec, the quality factor behaves well - the distribution is centered on the expected value of 1. At higher fluxes, the pixel response begins to be nonlinear. As the deviation from linearity increases, the quality factor increases as well, indicating that the straight line fit to the ramp is not correct any more. One should correct the fitted fluxes, as well as the quality factor for the nonlinearity effects. The method of the on-ground nonlinearity correction applied in Euclid will be presented in the future work.

Finally we have confirmed the sensitivity of the quality factor to the saturation, cosmic ray hits and electronic effects. The evaluation of the sensitivity to any of these effects as the function of the underlying flux, pixel readout noise as well as the amplitude of the anomaly is postponed to the next paper.

We are convinced that the proposed flux estimate and the associated quality factor will help achieve the high precision scientific goals of the mission. We anticipate that the method can be applied to other missions that use similar detectors and readout schemes and that are subject to similar CPU and telemetry limitations.

\acknowledgments 
The authors would like to thank the engineers and researchers form the Institut de Physique Nucl\'eaire de Lyon, from the Centre de Physique des Particules de Marseille and from the N\'eel Institute in Grenoble that have contributed to the campaigns of data taking for their efforts. The authors thank also the support from the CNRS/IN2P3. This research was conducted within the framework of the Lyon Institute of Origins under grant ANR-10-LABX-66.

\bibliography{Allrefs} 
\bibliographystyle{spiebib} 

\end{document}